\documentclass[a4paper,11pt]{article}
\usepackage{pos}
\usepackage{array}

\title{Cluster-Algorithm-Amenable Models of Gauge Fields and Matter}

\author*{Emilie Huffman\footnote{In collaboration with Debasish Banerjee}}

\affiliation{Perimeter Institute for Theoretical Physics,Waterloo, ON N2L 2Y5, Canada}

\emailAdd{ehuffman@perimeterinstitute.ca}

\abstract{Typical fermion algorithms require 
 the computation (or sampling) of the fermion determinant. We focus instead on cluster algorithms 
 which do not involve the determinant and involve a more physically relevant sampling of the configuration space. 
 We develop new cluster algorithms and design classes of models for fermions coupled to $\mathbb{Z}_2$ and $U(1)$ 
 gauge fields that are amenable to being simulated by these cluster algorithms in a sign-problem free way. Such simulations should contain rich phase diagrams and are particularly relevant for quantum simulator 
 experiments.}

\FullConference{The 40th International Symposium on Lattice Field Theory (Lattice 2023)\\
July 31st - August 4th, 2023\\
Fermi National Accelerator Laboratory\\}

\begin{document}
\maketitle

\section{Introduction}
Fermions that strongly interact with each other via gauge fields are an essential ingredient for many models in particle physics. For example, the fermions may constitute the matter component, such as leptons in 
electromagnetic and weak interactions, or quarks in strong interactions, and then the gauge fields may constitute
the photon, the $W^{\pm},Z$, and the gluons respectively. Gauge fields are also becoming
increasingly important to condensed matter systems, from frustrated magnetism to theories of deconfined 
quantum criticality. It thus is interesting to develop efficient simulations for interacting theories of gauge fields and matter.


  While Quantum Monte Carlo (QMC) methods are robust simulation methods for non-perturbative studies of the aforementioned systems, 
they are also vulnerable to the sign problem \cite{Troyer:2004ge}. QMC methods work by performing 
importance sampling of configurations that make up the partition function. Since fermions anti-commute, their sign problem can be straightforwardly understood when the configurations considered are 
worldlines: whenever fermions exchange positions an odd number of times, the configuration weight acquires another 
negative sign factor, leading to huge cancellations in the summation and an exponential scaling in the volume for calculations. 

Meron cluster methods \cite{Bietenholz_1995}, so named due to the presence of \textit{merons} (half-instantons)
in the first model for which they were developed to simulate (the 2d $O(3)$ sigma model with $\theta=\pi$), 
can solve sign-problems in 
four-fermion Hamiltonians for certain parameter regimes, as well as for free fermions with a chemical potential
\cite{Chandrasekharan:1999cm}. Because these methods sample worldlines, computing 
the weights scales linearly with the volume of the system, and negative terms in the partition function are taken 
care of by avoiding \textit{merons} --- this is what distinguishes them from bosonic simulations. Because there are none of the stabilization 
issues that can arise in determinantal methods, and the weight 
computations scale favorably compared to determinantal methods, these cluster methods are an attractive choice for simulation when applicable. Correspondingly, exciting opportunities 
open up when new interesting physical models are found which can be simulated using this method.

  Recently, there has been experimental development regarding the physics of confinement and quantum spin 
liquids using quantum simulation. The models used to capture the 
physics involve fermions interacting with (Abelian) gauge fields. In these proceedings we introduce new cluster algorithms and develop designer\cite{Kaul_2013} classes of experimentally relevant models which are sign-problem-free for these cluster simulations, enabling 
a robust elucidation of their phase diagrams. More details may be found in our full manuscript \cite{HufBan}. Moreover, the worldline nature of the configurations makes it easy to produce measurement ``snapshot'' synthetic data to compare with quantum simulators, and 
such configurations are promising inputs for machine learning algorithms.\cite{Bohrdt}

\begin{figure}[t]
    \centering
    \includegraphics[width=2.2in]{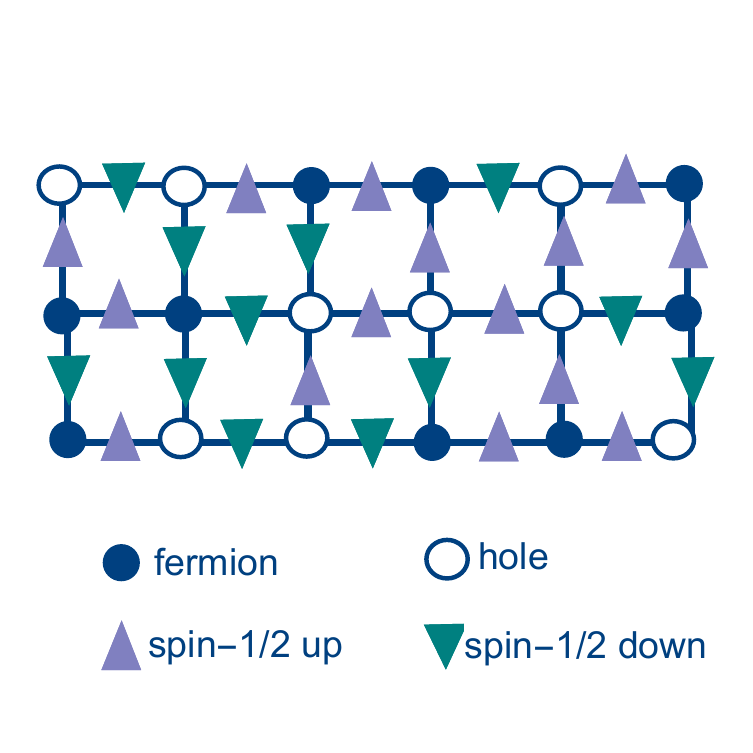}
    \caption{The lattice for these models of gauge fields and matter with $N_f=1$. The spin-$1/2$ degrees of freedom are on the links while the fermions may hop from one site to another.}
    \label{fig:lattice}
\end{figure}

\section{Models}
We build these cluster-algorithm amenable models by beginning with the half-filled $t$-$V$ model,
\begin{equation}
H = \sum_{\left\langle xy\right\rangle}\left[ -\frac{t}{2} \left(c^\dagger_x c_y + c^\dagger_y c_x\right) 
 + V \left(n_x - \frac{1}{2}\right) \left(n_y - \frac{1}{2}\right) \right].
\end{equation} 
 Here $\left\langle xy\right\rangle$ are nearest neighbor sites, $c^\dagger$ and $c$ are 
 creation and annihilation operators respectively, and the repulsive interaction $V$ is given in terms of the occupation number $n = c^\dagger c$. We begin with this model because it is simulable by meron clusters for $V\geq 2t$ 
 \cite{Chandrasekharan:1999cm, Liu:2020ygc}. 
 
 With this starting point, we now build up 
 cluster-amenable Hamiltonians involving gauge fields. They are organized into $\mathbb{Z}_2$- and $U(1)$-
gauge symmetric families, which involve the following terms:
\begin{equation}
    H^{(g)}_{N_f} = -\sum_{\left\langle xy\right\rangle} \prod_{f=1}^{N_f} \left(H^{(g)}_{\left\langle xy\right \rangle, f} + H^{(g),\mathrm{des}}_{\left\langle xy\right\rangle, f}\right)
    \label{eq:modelfam}
\end{equation}
The label $g\in\left\{U(1),\mathbb{Z}_2\right\}$ is the gauge symmetry, with
\begin{equation}
\begin{aligned}
    H^{\mathbb{Z}_2}_{\left\langle xy\right\rangle, f} &=  t\left(c^\dagger_{x,f}s^1_{xy,f}c_{y,f} +c^\dagger_{y,f}s^1_{xy,f}c_{x,f}\right),\\
    H^{U(1)}_{\left\langle xy\right\rangle, f}& = t\left(c^\dagger_{x,f}s^+_{xy,f}c_{y,f} +c^\dagger_{y,f}s^-_{xy,f}c_{x,f}\right).
    \end{aligned}
    \label{eq:local}
\end{equation}
The hopping of spinless fermions between the nearest neighbor sites $xy$ are now governed by the presence
of gauge fields, represented by spin-1/2 operators, $s^k_{xy}$, on the bond. Figure \ref{fig:lattice} illustrates the lattice with the gauge field and matter degrees of freedom. These terms are lower-dimensional versions of quantum electrodynamics 
 (QED). Then $H^{(g),\mathrm{des}}_{\left\langle xy\right\rangle, f}$ is a designer term\cite{Kaul_2013} that makes the models particularly 
amenable to the cluster algorithms,
(we consider $V = 2t$, for simplicity)
\begin{equation}
    \begin{aligned}
H^{\mathbb{Z}_2,\mathrm{des}}_{\left\langle xy\right\rangle, f} &= -2t\left(n_{x,f}-\frac{1}{2}\right)\left(n_{y,f}-\frac{1}{2}\right)+ \frac{t}{2}\\
    H^{U(1),\mathrm{des}}_{\left\langle xy\right\rangle, f} &= -t\left(n_{x,f}-\frac{1}{2}\right)\left(n_{y,f}-\frac{1}{2}\right)-t s_{xy,f}^3\left(n_{y,f}-n_{x,f}\right) + \frac{t}{4}
    \end{aligned}
    \label{eq:design}
\end{equation}
\begin{figure}[t]
    \centering
    \includegraphics[width=1.1in]{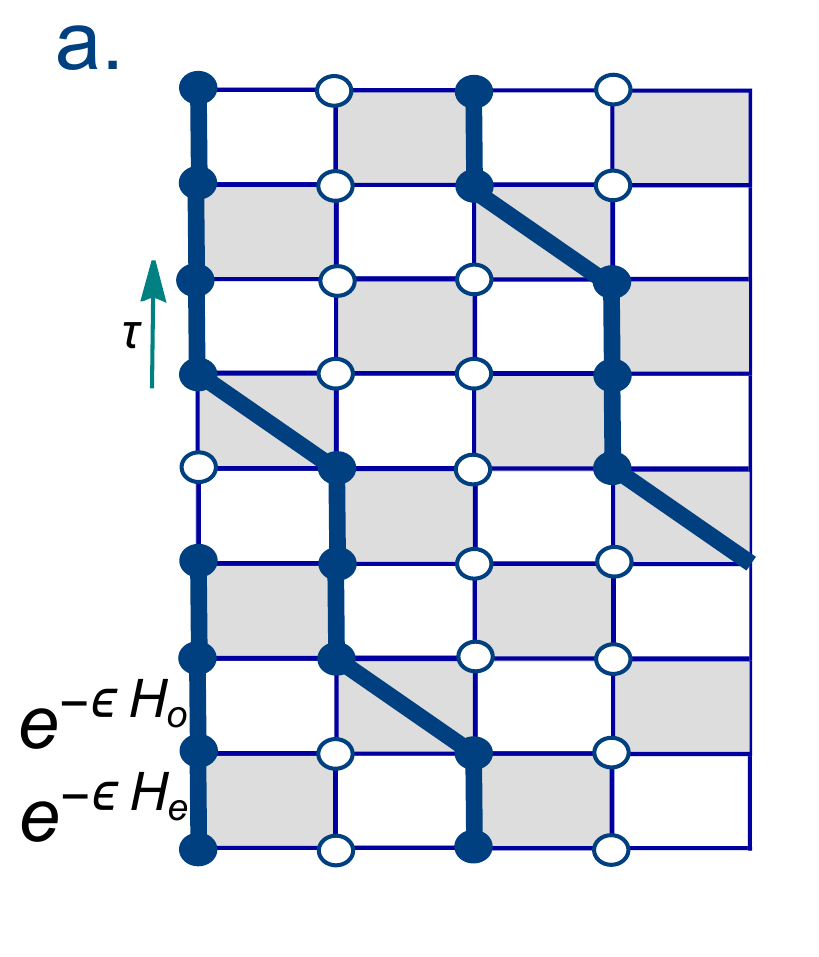}
    \includegraphics[width=1.1in]{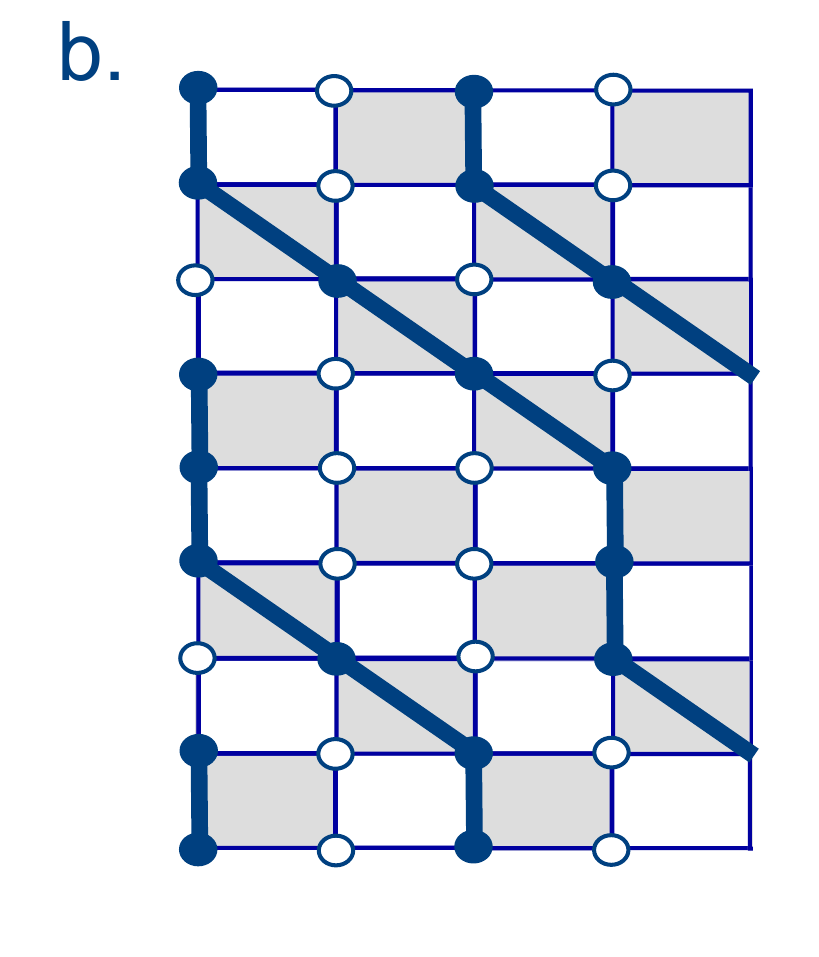}
    \includegraphics[width=1.1in]{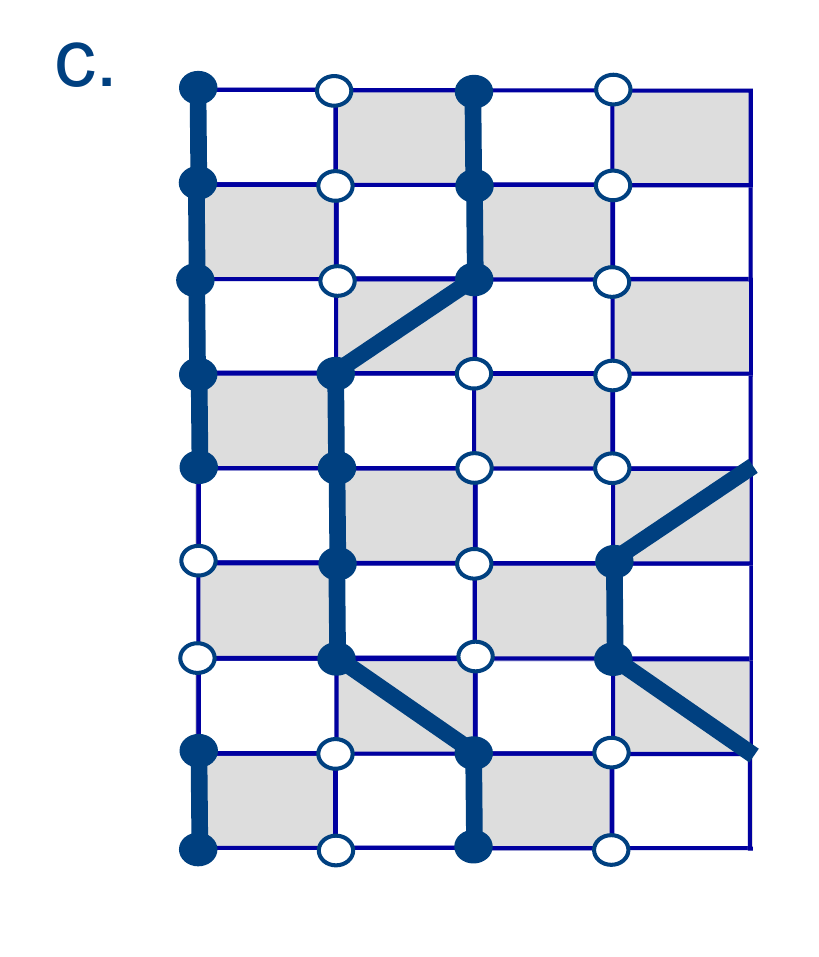}
    \includegraphics[width=1.1in]{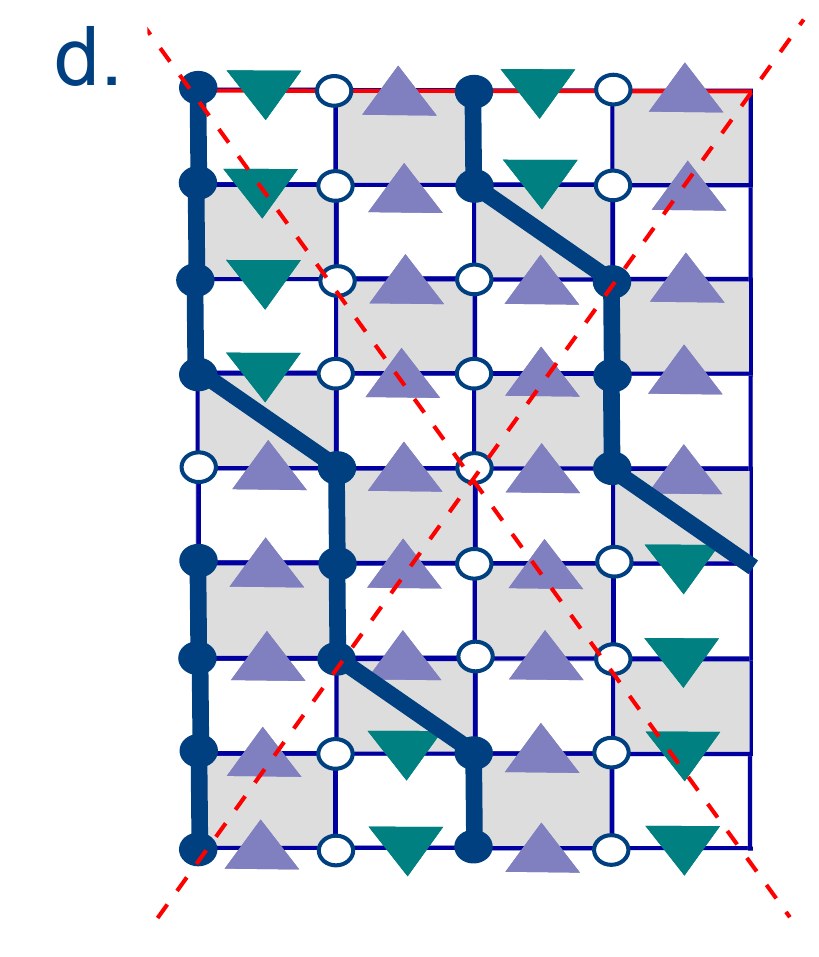}
    \includegraphics[width=1.1in]{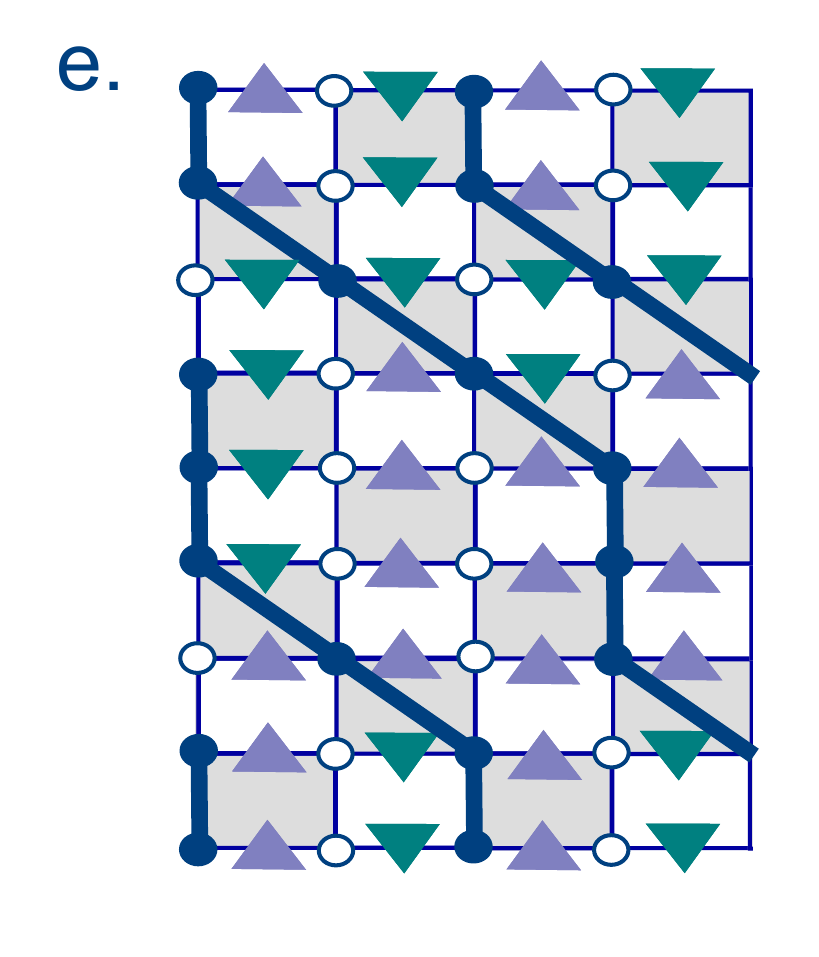}
    \includegraphics[width=1.1in]{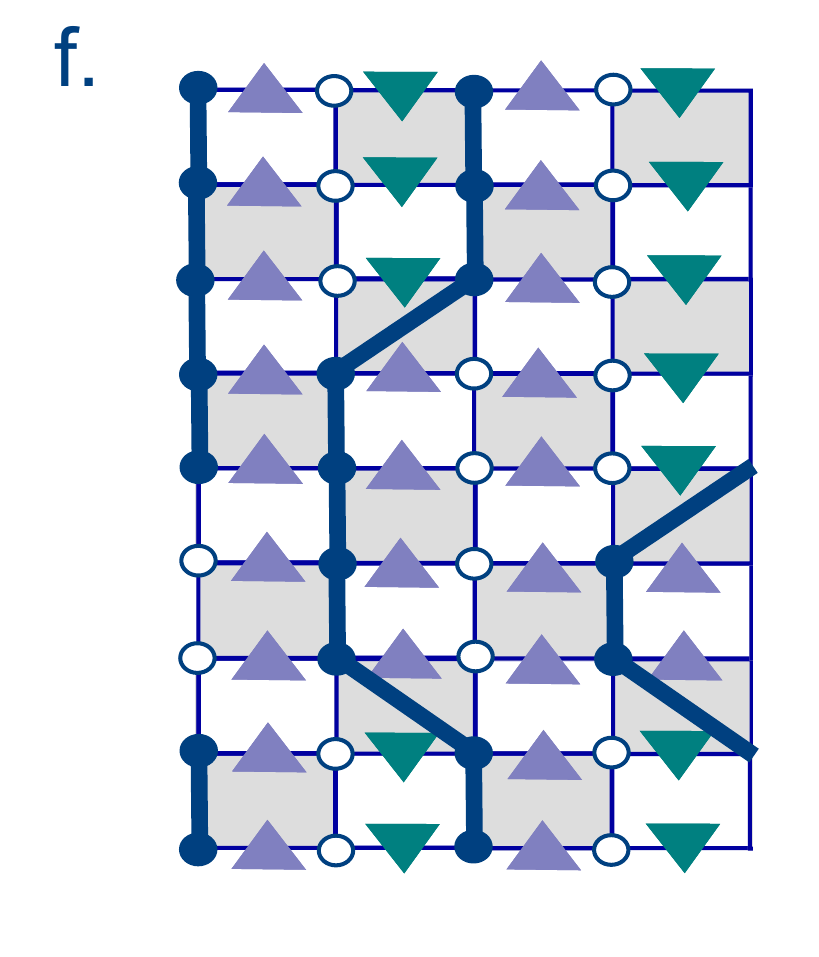}
    \includegraphics[width=1.1in]{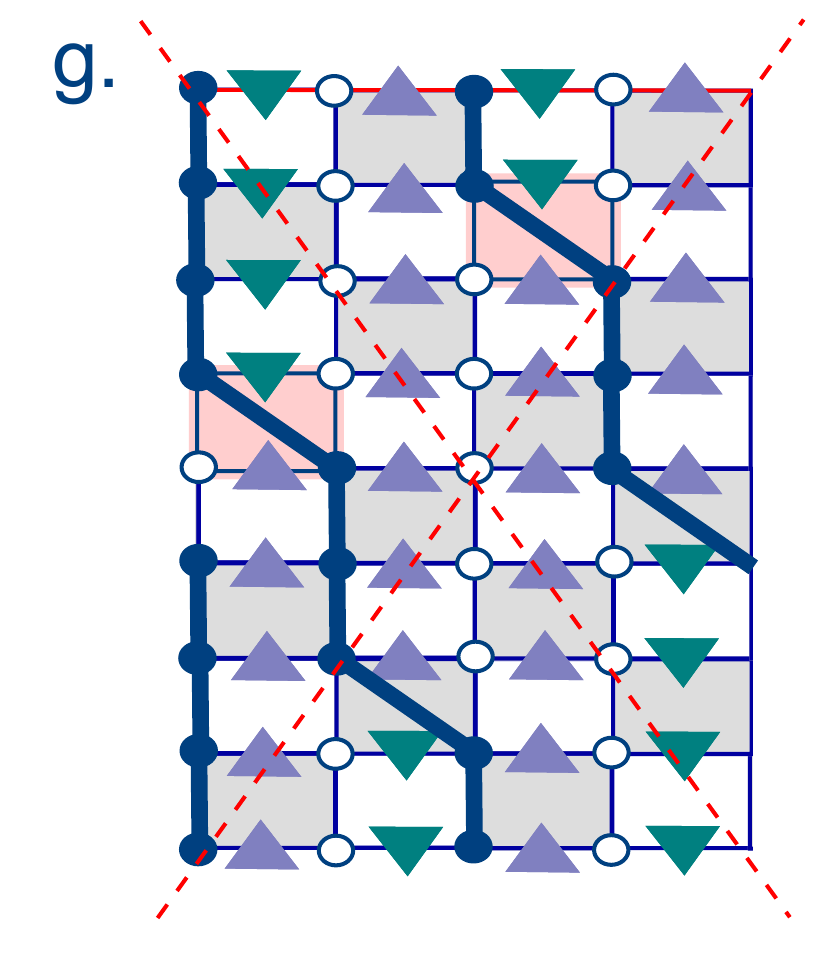}
    \includegraphics[width=1.1in]{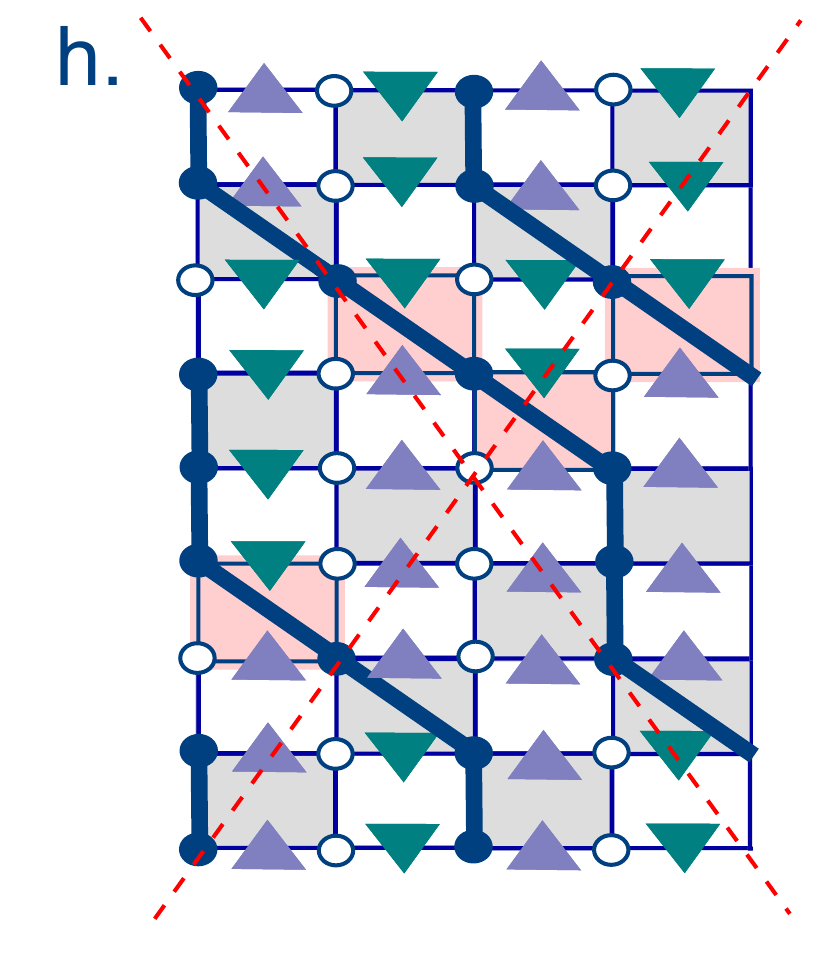}
    \includegraphics[width=1.1in]{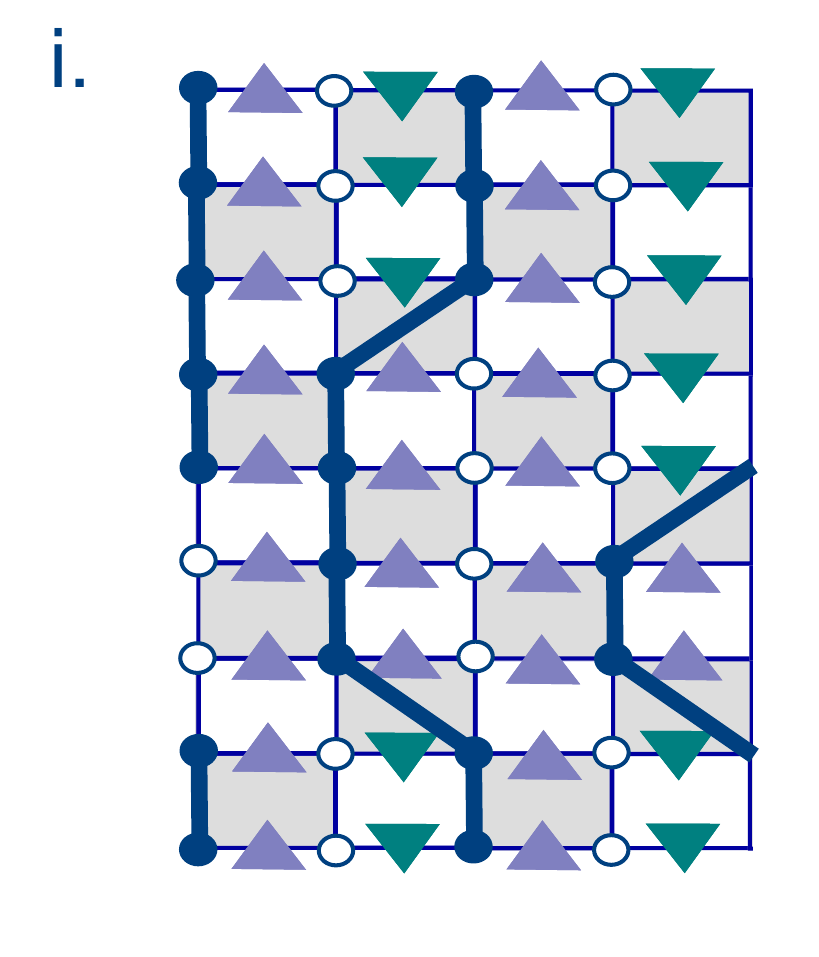}
    \caption{Worldlines for the $t$-$V$ model are in (a)-(c), the $\mathbb{Z}_2$ theory in (d)-(f), and the $U(1)$ 
    theory in (g)-(i). Image (a) shows the imaginary time direction and the $(1+1)$-d trotterization, which is the same for all images. Filled circles are sites occupied by fermions, and empty circles are holes. Figures in the second two rows also have link variables because they correspond to gauge theories: the upward triangles correspond to spin $+1/2$ and the downward triangles correspond to spin $-1/2$. While the fermionic worldlines are the same in each column, some configurations that are allowed for the $t$-$V$ model have zero weight for the $\mathbb{Z}_2$ and $U(1)$ theories. These are crossed out, and zero-weight plaquettes are shaded red.}
    \label{fig:worldlines}
\end{figure}

For the $\mathbb{Z}_2$ gauge 
theory, the Pauli matrix $s^1=\sigma^1/2$ couples to fermions, and one can see the local $\mathbb{Z}_2$ symmetry 
from the operator $Q_x$, which commutes with the $H^{\mathbb{Z}_2}$ and is given by 
$Q_x = (-1)^{\sum_f n_{x,f}}\prod_{f,\hat{\alpha}} s^3_{x,x+\hat{\alpha},f}  s^3_{x-\hat{\alpha},x,f}$, 
where $\hat{\alpha}$ are the unit vectors in a $d$-dimensional square lattice. For the $U(1)$ theory, one can see 
the $U(1)$ symmetry from the unitary operator $V_{U(1)}$ which commutes with $H^{U(1)}$ and is given by 
$V_{U(1)} = \prod_x e^{i\theta_x G_x}$, with $G_x = \sum_{f} \left[n_{x,f} - 
    \sum_{\hat{\alpha}}\left(s^3_{x,x+\hat{\alpha},f}- s^3_{x-\hat{\alpha},x,f}\right)+((-1)^x-1)/2\right]$.
 These models are \emph{quantum link models} \cite{Wiese:2021djl}, 
which realize the continuous gauge invariance using finite-dimensional quantum degrees of freedom. The identification 
with usual gauge field operators is given by $U_{xy,f} = s^+_{xy,f},~ U^\dagger_{xy,f} = s^-_{xy,f}, E = s^3_{xy,f}$.
 We note that a straightforward application of the meron idea necessitates the introduction of an equivalent 
\emph{flavor} index for gauge links as fermion flavors. The total Gauss law can be expressed through a 
product ($\mathbb{Z}_2$) or sum ($U(1)$) of the Gauss law of individual flavors degrees of freedom, and the resulting 
theories have $\mathbb{Z}_2^{\otimes N_f}$ and $U(1)^{\otimes N_f}$ gauge symmetry. However, flavored gauge-interactions 
can also be turned on in the $U(1)$ model (as explained in our manuscript \cite{HufBan}), $H^{U\left(1\right)}_{N_f=2} \rightarrow H^{U\left(1\right)}_{N_f=2} + J\sum_{\left\langle xy\right\rangle} s^3_{xy,1} s^3_{xy,2}$, or through a Hubbard-U interaction for both $\mathbb{Z}_2$ and $U(1)$-symmetric models \cite{Liu:2020ygc}. 
These additions would directly cause ordering for either the gauge fields or fermions, with the coupling 
between them leading to the interesting question of how the other degrees of freedom are affected by this ordering.

\section{Algorithm}
We first consider the possible worldline configurations for the base models 
 defined in (\ref{eq:modelfam}) and (\ref{eq:local}). Here we use the occupation number basis for the 
 fermions and the electric flux (spin-$z$) basis. The partition function in $(1+1)$-d is then
\begin{equation}
\begin{aligned}
    &{\cal Z} ={\mathrm{Tr}} \left(e^{-\beta H}\right) = \sum_{\left\{s,n\right\}} \left\langle s_1,n_1\right| e^{-\epsilon H_e}\left|s_{2N_t},n_{2N_t}\right\rangle\left\langle s_{2N_t},n_{2N_t}\right|\\ &\qquad\qquad\qquad\qquad\qquad\qquad\times e^{-\epsilon H_o}...e^{-\epsilon H_e}\left|s_2,n_2\right\rangle\left\langle s_2,n_2\right|
    e^{-\epsilon H_o}\left|s_1,n_1\right\rangle ,
    \end{aligned}
    \label{eq:worldlines2}
\end{equation}
where $H = H_e + H_o$, and $H_e$ ($H_o$) consists of Hamiltonian terms that correspond to even (odd) links. 
This is a Trotter approximation, and all terms within $H_e$ and $H_o$ commute with each other. We thus have a sum of terms 
that consist of discrete time-slices $1, \cdots, 2N_t$, with defined electric flux and fermion occupation 
numbers for each of the time-slices. Each of the terms in (\ref{eq:worldlines2}) is a worldline configuration. Figure \ref{fig:worldlines}(a)-(c) give examples of such configurations for the $t$-$V$ 
model as simulated by meron clusters. We note that these configurations are similar to the ``snapshots'' obtained from experiment in \cite{Bohrdt}, but involve more information from the additional imaginary time direction (one snapshot would be one horizontal line in a configuration from Figure \ref{fig:worldlines}), so we are able to obtain imaginary time displaced correlations from these configurations.

In the $\mathbb{Z}_2$ case, for each time-slice a fermion may hop to an unoccupied nearest neighbor site of the same flavor. If the fermion hops, then the electric flux on the bond between the nearest neighbor sites that shares the same fermion flavor index must also flip--this is due to the $s_{xy}^1$ operator. Figure \ref{fig:worldlines}(d)-(f) gives example configurations for the $N_f=1$ version of 
this model. Due to the trace, it is impossible to have odd winding numbers because these would cause the spins in the initial state to not match the spins in the final state. The possible worldline configurations for the $U(1)$ are similar to the $\mathbb{Z}_2$ case, but more restrictive. 
The $s^+_{xy}$ and $s^-_{xy}$ operators allow the hopping for a given flavor of fermion only in a single direction for each bond, which depends on the orientation of the same flavored flux on the bond. 
Figure \ref{fig:worldlines}(g)-(i) illustrates an example configuration and restrictions for the single flavor version of 
the $U(1)$ model. In $(1+1)$-d it is clear that all allowed configurations must 
have zero winding number.


The worldline configurations are a tool to obtain cluster configurations by introducing appropriate breakups, 
which decompose the terms in (\ref{eq:worldlines2}) into further constituents. In considering the allowed worldline 
configurations given in (\ref{fig:worldlines}) for the $U(1)$ theory, for example, each of the active plaquettes in each 
time-slice (shaded in gray) must be one of the plaquettes given in Table \ref{tab:first}. The plaquettes in each row 
share the same weight, computed using $\left\langle s_b, n_b\right| e^{-\epsilon H_b}\left|s_b', n_b'\right\rangle$, 
from (\ref{eq:worldlines2}), where $b$ is a nearest neighbor bond, $b=\{x,y\}$. The corresponding breakup cell for each 
row gives allowable breakups: if all fermion occupations/spins are flipped along any one of the lines, the resulting 
plaquette also exists in this table. From the table, we see two such breakups are defined, $A$ and $D$. These breakups resemble the breakups from the original fermionic meron cluster algorithms, but also involve the link variables--either as additional lines for the $A$ breakups, or as \textit{binding} lines extending outward from the horizontal $D$ breakup lines. This is a key difference for the new gauge version of the algorithm. By computing 
the matrix elements for the plaquettes in each grouping, we find that for the $U(1)$ theory, the 
corresponding breakup weights $w_A$ and $w_D$ must obey:
\begin{equation}
    \begin{aligned}
    w_A &= 1 \\
    w_D &= \exp \left(\epsilon t\right) \sinh \epsilon t \\
    w_A + w_D &= \exp\left(\epsilon t\right) \cosh \epsilon t ,
    \end{aligned}
\end{equation}
\begin{table}
    \centering
    \begin{tabular}{ m{2.7cm} | m{2.9cm}}
    \hline
        Plaquettes & Breakups  \\
         \hline
         \includegraphics[width=1.3cm]{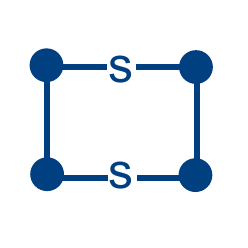} 
         \includegraphics[width=1.3cm]{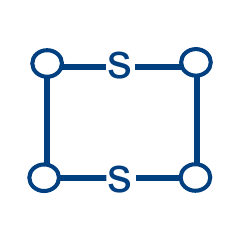} 
         \includegraphics[width=1.3cm]{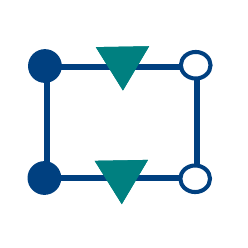} 
         \includegraphics[width=1.3cm]{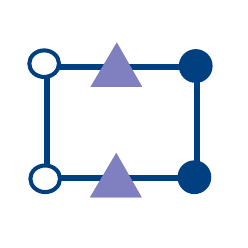}   &
         
         \includegraphics[width=1.3cm]{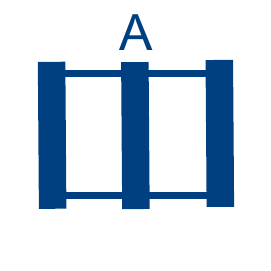}
         \includegraphics[width=1.3cm]{Paper/Figures/blank.pdf}
         \includegraphics[width=1.3cm]{Paper/Figures/blank.pdf}
         \includegraphics[width=1.3cm]{Paper/Figures/blank.pdf}
         \\
         \hline
         \includegraphics[width=1.3cm]{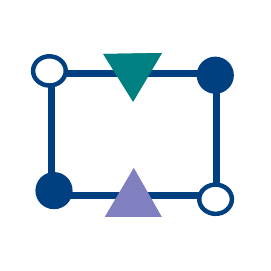}\includegraphics[width=1.3cm]{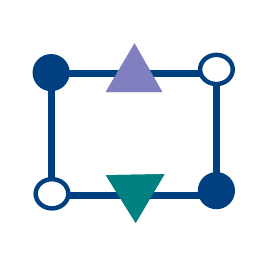} &
         \vspace{.1cm}
         \includegraphics[width=1.3cm]{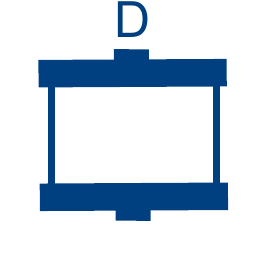} \\
         \hline
         \includegraphics[width=1.3cm]{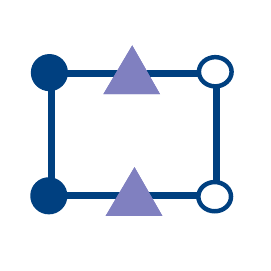}\includegraphics[width=1.3cm]{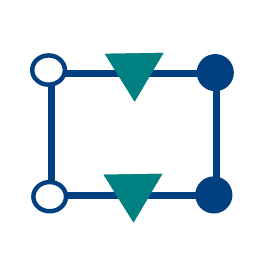} &
         \vspace{0.05cm}
         \includegraphics[width=1.3cm]{Paper/Figures/pa.pdf}  \includegraphics[width=1.3cm]{Paper/Figures/dbreakup2.pdf}    
         \vspace{.05cm}\\
         \hline
    \end{tabular}
    \caption{Plaquettes and breakups for the $U(1)$-symmetric Hamiltonian. The middle vertical cluster lines on the $A$-breakups and gauge binding lines on the $D$-breakups distinguish them from the purely fermionic versions of these breakups.}
    \label{tab:first}
\end{table}
to satisfy detailed balance. We can then simulate this system similarly to the simulation of the $t$-$V$ model at $V=2t$
\cite{Chandrasekharan:1999cm}, by exploring a configuration space where each configuration is defined by the worldlines and breakups. By assigning breakups to all active plaquettes, we form clusters, and then updates involve flipping all fluxes and fermions within a cluster, which physically corresponds to generating a new 
worldline configuration. The algorithm begins by putting the system in a \textit{reference configuration}, defined by the fermionic worldlines only, where 
the weight is known to be positive. It is straightforward to see that all configurations can be returned 
to the fermion worldlines of this configuration by cluster flips. For 
both the $U(1)$ and $\mathbb{Z}_2$ theories, the reference configuration has a staggered fermionic occupation, where fermions are stationary throughout imaginary time. Fluxes and breakups may be initially attached to spacetime plaquettes in any nonzero configuration according to Table \ref{tab:first}. A QMC sweep is then:
\begin{enumerate}
    \item Go through the list of the active plaquettes and update each breakup sequentially.
    \begin{enumerate}
        \item If the breakup can be changed for a plaquette, change it with probability dependent on the breakup weights.
        \item If the breakup is changed, consider the new configuration that would result. If it contains 
        a cluster where flipping the fermion occupation causes the fermions to permute in a way that produces a negative 
        sign, then it is a \textit{meron}. In that case, change the breakup back to its initial state. Rules for 
        identifying merons generalize \cite{Chandrasekharan:1999cm} and are given in greater detail below.
    \end{enumerate}
    \item Identify the new clusters from the breakups in the new configuration. For each cluster, flip all 
    fermion occupation states and spins with probability $1/2$.
\end{enumerate}
This describes sampling of the zero-meron sector only, but sectors with other numbers of merons are relevant for off-diagonal observables. \cite{Chandrasekharan:1999cm} While these cluster rules implement the Hamiltonian 
dynamics, Gauss law constraints are not included.

\begin{figure}
    \centering
    \includegraphics[scale=0.185]{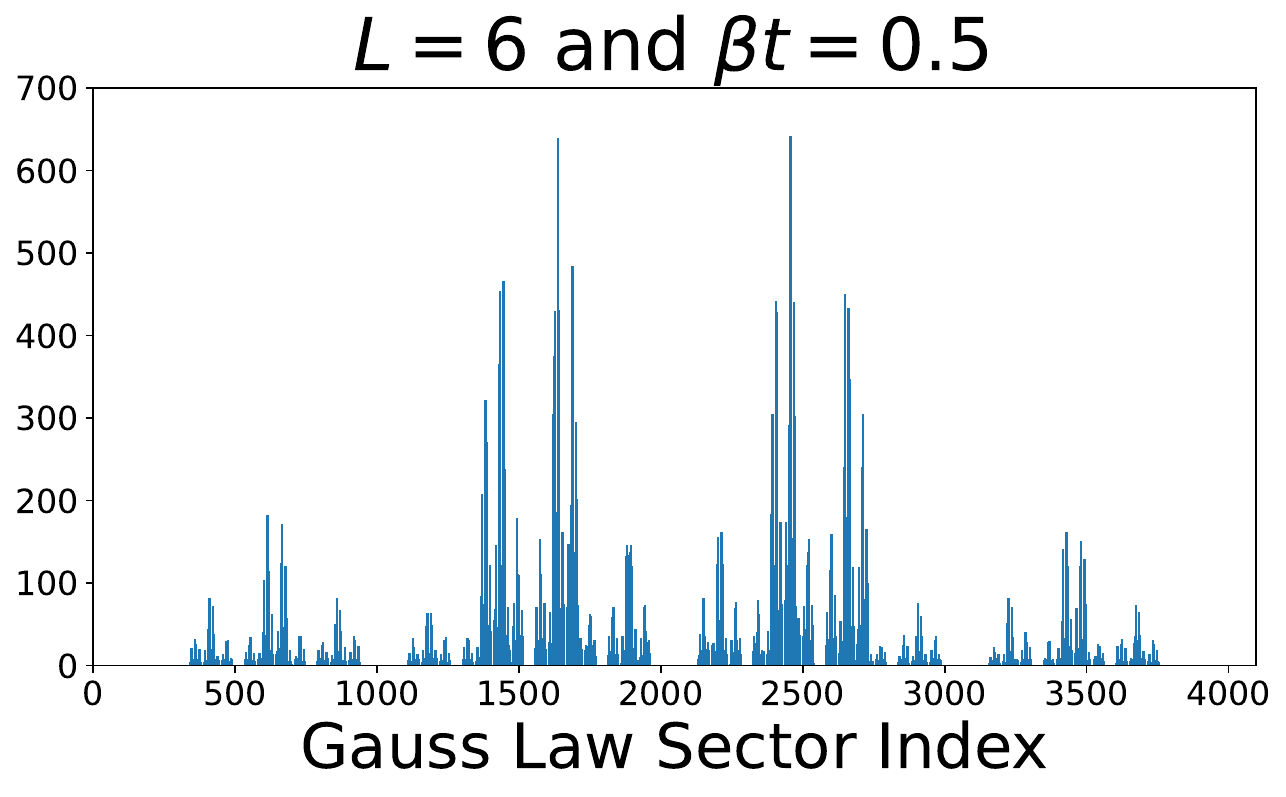}
    \includegraphics[scale=0.185]{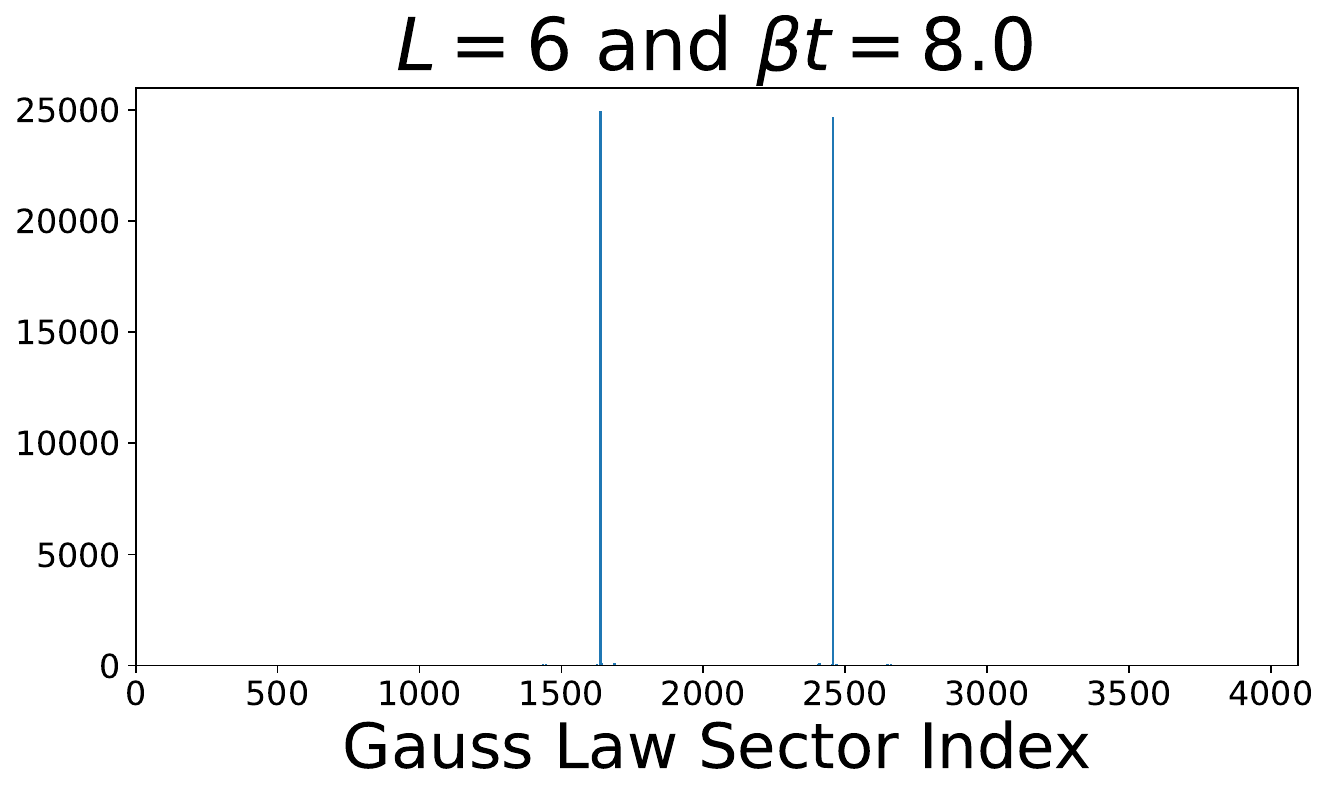}
    \caption{Number of configurations versus Gauss law sector index $\sum_x\left[G_x+2\right] \cdot 4^x$ (not all indices correspond to actual sectors) for $50000$ equilibrated 
    configurations. Two sectors emerge at large $\beta$: $G_x = 0$ and $G_x = (-1)^x$. 
    }
    \label{fig:GaussLaw}
\end{figure}
\section{Results}
To illustrate the efficacy of the algorithm, we discuss results obtained by simulating the 
$(1+1)$-d $H^{U(1)}_{N_f=1}$ model in (\ref{eq:modelfam}), which is related to the massless 
quantum-link Schwinger model and the PXP model, where quantum scars were first demonstrated experimentally 
\cite{Bernien_2017}. We simulate the model for different temperatures $\beta=1/T$, 
\emph{without imposing the Gauss law}. A filter may then be applied to study the physics in the desired Gauss law sector. The first non-trivial result is the 
emergence of \emph{two} Gauss' law sectors at low temperatures, as shown in Figure \ref{fig:GaussLaw}, other results are given in our manuscript \cite{HufBan}. 

\begin{figure}
\includegraphics[width=3.5cm]{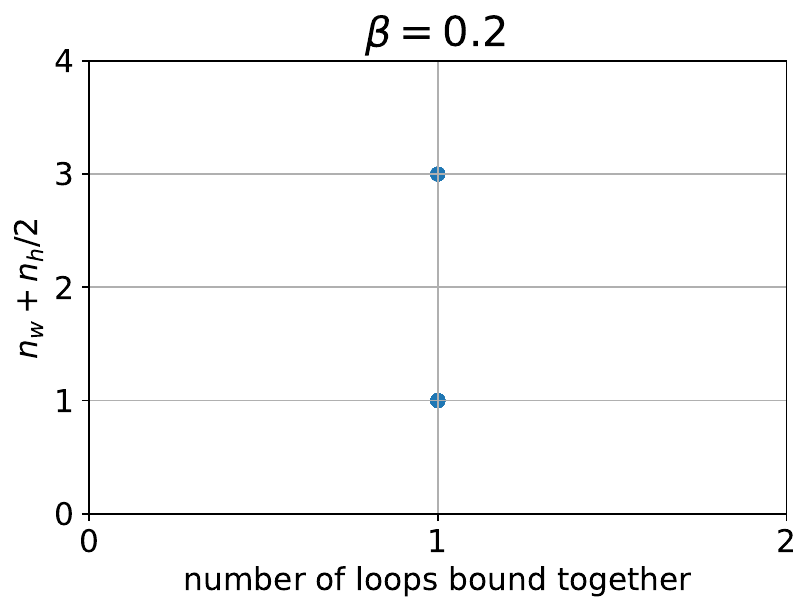}
\includegraphics[width=3.5cm]{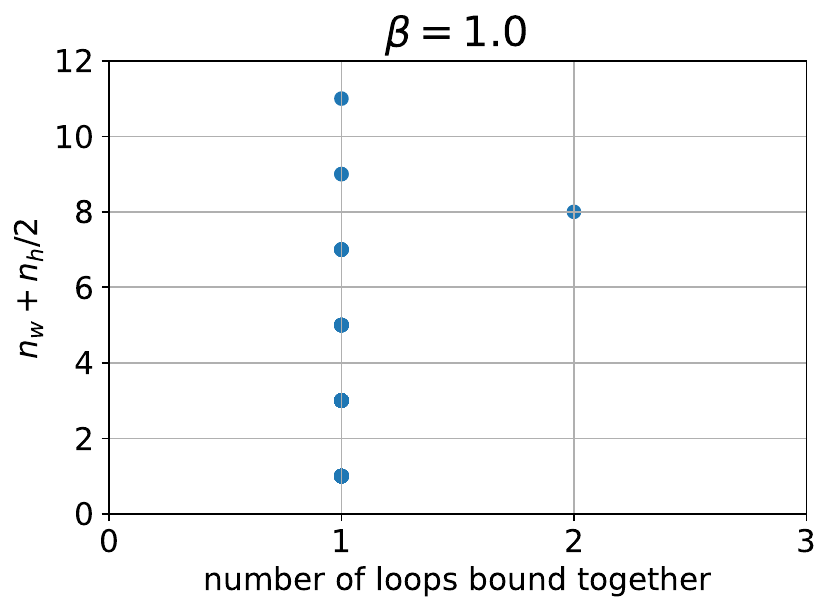}
\includegraphics[width=3.5cm]{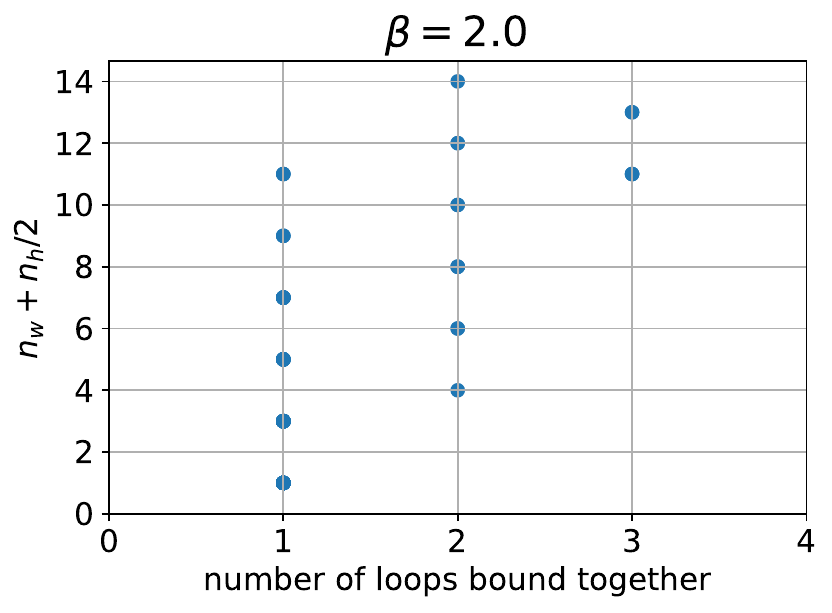}
\includegraphics[width=3.5cm]{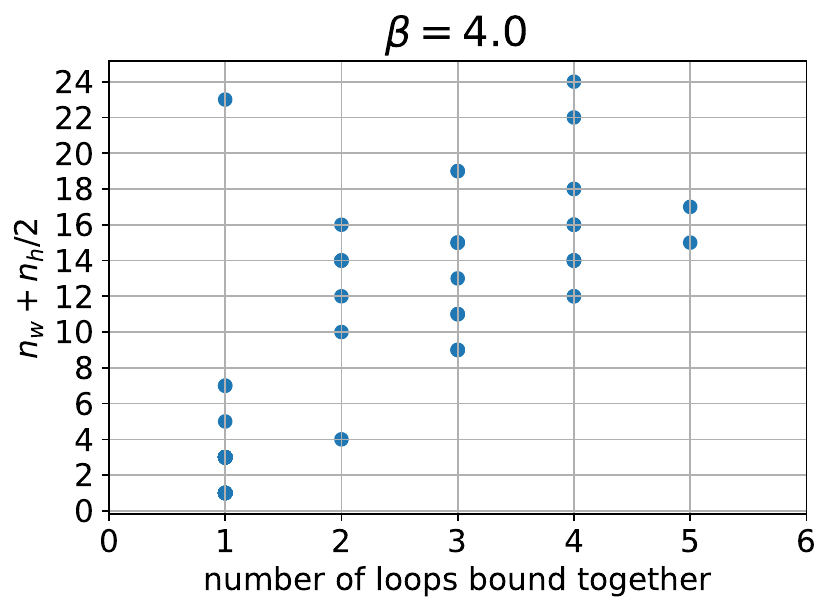}
\caption{The types of clusters that appear as a function of temperature. None of them are merons because the simulation is in $1+1$d.}
\label{clusters}
\end{figure}

For purely fermionic models, as shown in \cite{Chandrasekharan:1999cm}, a loop is a meron if the 
quantity $n_w + n_h/2$ is even, where $n_w$ is the number of temporal windings and $n_h$ is the 
number of fermionic hops in the loop. For the new gauge-fermion algorithm, we now can bind two or more loops to each other through the gauge fields, and thus we have the 
following new meron criterion for a cluster
\begin{equation}
\mathrm{Meron \; if} \left\{\begin{array}{cc} n_w + n_h/2\; \mathrm{odd}, & \mathrm{even\;} \# \mathrm{\;of \;loops}\\
n_w + n_h/2\; \mathrm{even}, & \mathrm{odd\;} \# \mathrm{\;of \;loops}\end{array}\right.
\label{meroncriterion}
\end{equation}
It can be seen immediately that (\ref{meroncriterion}) reduces to the original definition in the case 
of one loop. In one dimension there should be no merons. Figure 
\ref{clusters} plots the $n_h$ and $n_w$ for $100$ equilibrated clusters at low temperature, and 
indeed we see that by the criterion, none of these clusters are merons.
\section{Conclusions}
We have generalized the construction of the meron algorithm to cases where staggered fermions are
coupled to quantum link gauge fields. This construction of the Monte Carlo algorithm is agnostic to the space-time 
dimension, and paves the way for unbiased studies of large scale gauge-fermionic system with odd or even numbers 
of fermionic flavours, and includes models not simulable using determinantal Quantum Monte Carlo. It is possible to add different microscopic 
terms by increasing the allowed ways of bonding the fermions and gauge links. Future extensions could include gauge fields with larger spin representation and non-Abelian gauge fields as well.

\begin{thebibliography}{99}

\bibitem{Troyer:2004ge}
Matthias Troyer and Uwe-Jens Wiese (1994) \emph{Computational complexity and fundamental limitations to fermionic quantum Monte Carlo simulations.}, Phys. Rev. Lett., 94:170201, 2005.

\bibitem{Bietenholz_1995}
W. Bietenholz, A. Pochinsky, and U. J. Wiese. \emph{Meron cluster simulation of the $\theta$ vacuum in the 2d o(3) model.}
Physical Review Letters, 75(24):4524–4527, Dec 1995.

\bibitem{Chandrasekharan:1999cm}
Shailesh Chandrasekharan and Uwe-Jens Wiese. \emph{Meron cluster solution of a fermion sign problem.} Phys. Rev. Lett., 83:3116–3119, 1999

\bibitem{Kaul_2013}
Ribhu K. Kaul, Roger G. Melko, and Anders W. Sandvik. \emph{Bridging lattice-scale physics and continuum field theory
with quantum monte carlo simulations.} Annual Review of Condensed Matter Physics, 4(1):179–215, Apr 2013

\bibitem{HufBan}
Debasish Banerjee and Emilie Huffman. \emph{Quantum Monte Carlo for Gauge Fields and Matter without the Fermion Determinant
} arXiv:2305.08917, 2023.

\bibitem{Bohrdt}
Annabelle Bohrdt, and et al. \emph{Classifying Snapshots of the Doped Hubbard Model with Machine Learning} Nature Physics volume 15, pages921–924 (2019)

\bibitem{Liu:2020ygc}
Hanqing Liu, Shailesh Chandrasekharan, and Ribhu K. Kaul. \emph{Hamiltonian models of lattice fermions solvable by the meron-cluster algorithm.} Phys. Rev. D,
103(5):054033, 2021.



\bibitem{Bernien_2017}
Hannes Bernien, and et al. \emph{Probing many-body dynamics on a
51-atom quantum simulator.} Nature, 551(7682):579–584,
Nov 2017.

\bibitem{Wiese:2021djl}
Uwe-Jens Wiese. \emph{From quantum link models to D-theory: a resource efficient framework for the quantum simulation
and computation of gauge theories.} Phil. Trans. A. Math. Phys. Eng. Sci., 380(2216):20210068, 2021.





\end{thebibliography}
\end{document}